# Rubidium transitions as wavelength reference for astronomical Doppler spectrographs

D. Rogozin*a,b, T. Fegerb, C. Schwabb , Y. V. Gurevichb, G. Raskina, D. W. Couttsb, J. Stuermerc, A. Seifahrtc, T. Fuehrer d, T. Legeroe, H. van Winckela, S. Halversonf, A. Quirrenbachg

aInstitute of Astronomy, KU Leuven, Celestijnenlaan 200D, 3001 Leuven, Belgium; bDepartment of Physics and Astronomy, Macquarie University, Macquarie Park, NSW 2109, Australia; cDepartment of Astronomy & Astrophysics, The University of Chicago, Chicago, IL 60637, United States; dInstitute of Applied Physics, Technische Universitaet Darmstadt, D-64289 Darmstadt, Germany; eDivision of Optics, Physikalisch-Technische Bundesanstalt, 38116 Braunschweig, Germany; fJet Propulsion Laboratory, Pasadena, CA 91109, United States; gLandessternwarte, ZAH, Koenigstuhl 12, 69117 Heidelberg, Germany

**ABSTRACT**

Precise wavelength calibration is a critical issue for high-resolution spectroscopic observations. The ideal calibration source should be able to provide a very stable and dense grid of evenly distributed spectral lines of constant intensity. A new method which satisfies all mentioned conditions has been developed by our group. The approach is to actively measure the exact position of a single spectral line of a Fabry-Perot etalon with very high precision with a wavelength-tuneable laser and compare it to an extremely stable wavelength standard. The ideal choice of standard is the D2 absorption line of Rubidium, which has been used as an optical frequency standard for decades. With this technique, the problem of stable wavelength calibration of spectrographs becomes a problem of how reliably we can measure and anchor one etalon line to the Rb transition. In this work we present our self-built module for Rb saturated absorption spectroscopy and discuss its stability.

## 1. INTRODUCTION AND EXPERIMENT

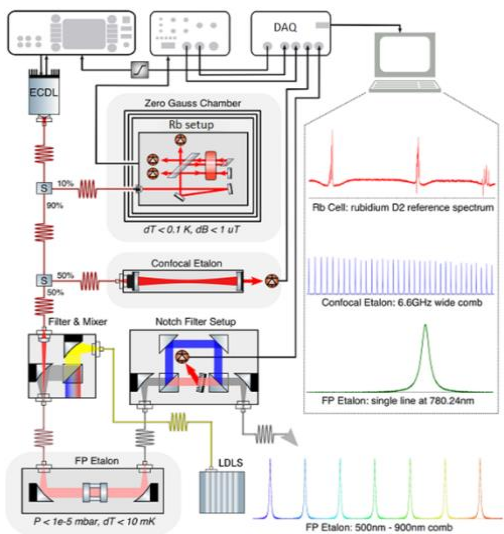

Figure 1. Schematic layout of the complete calibrator system.

Though there are well-established and widely used traditional methods for wavelength calibration of astronomical spectrographs like hollow-cathode lamps [1] and laser-frequency combs (LFC) [2] they all have some disadvantages. Lamps degrade with time, their emission lines are not evenly distributed along the spectrum and don`t have the same intensity. These lead to limitation in calibration precision one can achieve with them. In case of LFC the main disadvantage is their very high price, which makes it unaffordable for most of observatories.

New approaches that aim for ultimate stability and being not expensive were proposed. One of them is based on a Fabry-Perot etalon. But unlike in case with passively stabilised etalons [3] we intend for active tracing of one etalon line and locking it to the hyperfine transitions of the D2 absorption line of Rb [4]. The schematic layout of our complete calibrator system is shown in Figure 1 and consists of: external-cavity diode laser working at 780 nm, science Fabry-Perot etalon in vacuum chamber, Rb saturated absorption spectroscopy setup, confocal etalon for linearizing laser frequency scan and white light source.

The drawing of the Rb setup and the photo of the first built and assembled prototype can be found in Figure 2a & 2b. The probe beam is produced by retroreflecting the pump beam from an uncoated glass surface after it passes through the Rb cell and the beams are made to coincide by adjusting the kinematic mount. The saturated absorption is monitored via the signal photodiode. The monolithic bench to which all optics are mounted is to be temperature controlled and mounted kinematically inside a hermetic enclosure for ultra-stable operation.

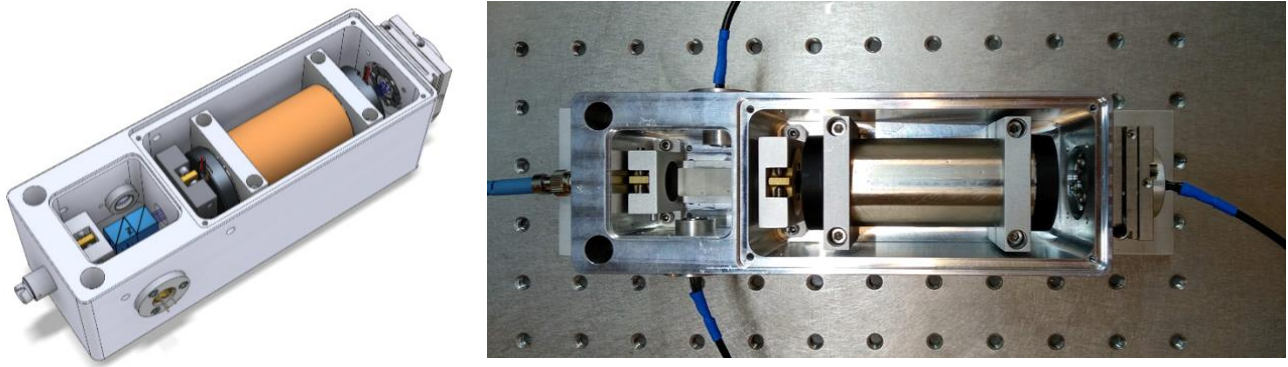

Figure 2. a) CAD rendering of the Rb spectroscopy setup; b) Photo of the first prototype of the Rb spectroscopy setup.

## 2. RESULTS AND DISCUSSION

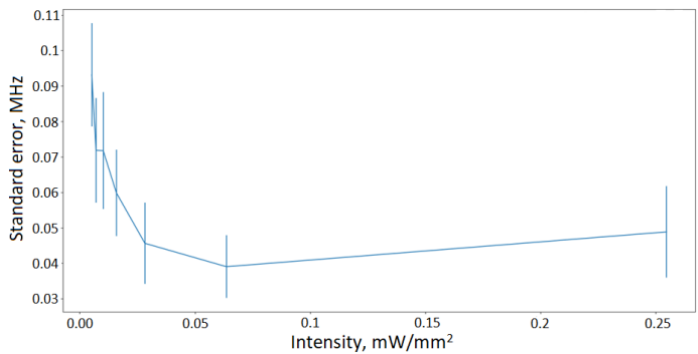

Figure 3. Dependency of the standard error of the line position obtained from the fit on the pump beam intensity.

In our calibrator system F2 transition of the Rb D2 line is exploited as a reference. It provides with 6 hyperfine lines and 4 of them are used in the analysis. The measured transition frequency is potentially affected by a number of factors: pump beam power, pump beam diameter, pump beam polarisation, pump-probe beam alignment, pump-probe beam intensity ratio, Rb cell temperature, ambient magnetic field etc. With long term stability of the setup as guiding principle we thoroughly study its sensitivity to the above mentioned factors.

As an example, in Figure 3 one can see dependency of the standard error of particular F2cF2F3 hyperfine line position obtained from the fit on the pump beam intensity, which was measured before the Rb cell. The error is below 0.1 MHz for all measured intensities and it corresponds to the radial velocity precision better than 10 cm/s. At the moment sensitivity of the Rb spectroscopy setup to other external factors is being investigated. Any improvements identified will be incorporated into a new version of the Rb setup.

*rogozin_d@ukr.net